\documentclass{article}
\usepackage{spconf,graphicx}
\usepackage{lineno,hyperref}
\usepackage{multirow}
\usepackage[utf8]{inputenc}
\usepackage{mathptmx}       
\usepackage{helvet}         
\usepackage{courier}        
\usepackage{type1cm}        
\usepackage{makeidx}         
\usepackage{multicol}        

\makeindex             
\usepackage[justification=centering]{caption}
\usepackage{graphicx}
\usepackage{url}
\usepackage{microtype}
\usepackage{tikz}
\usepackage{amsmath}
\title{Joint model-based recognition and localization of overlapped acoustic events using a set of distributed small microphone arrays}
\name{Rupayan Chakraborty and Climent Nadeu}
\address{TALP Research Center \\
Department of Signal Theory and Communications\\
Universitat Polit\`{e}cnica de Catalunya, Barcelona, Spain \\
\small \tt \{rupayan.chakraborty,climent.nadeu\}@upc.edu}
\begin{document}
\maketitle

\abstract{In the analysis of acoustic scenes, often the occurring sounds have to be detected in time, recognized,
and localized in space. Usually, each of these tasks is done separately. In this paper, a model-based
approach to jointly carry them out for the case of multiple simultaneous sources is presented and tested. The recognized event classes and 
their respective room positions are obtained with a 
single system that maximizes the combination of a large
set of scores, each one resulting from a different acoustic event model and a different beamformer output signal, which comes from one of 
several arbitrarily-located small microphone arrays.
By using a two-step method, the experimental work for a specific scenario consisting of meeting-room
acoustic events, either isolated or overlapped with speech, is reported. Tests carried out with two datasets show the advantage of the 
proposed approach with respect to some usual techniques, and that the inclusion of estimated priors brings a further performance improvement.}


\section{Introduction}
\label{sec:introduction}

The automatic analysis of acoustic scenes requires several functionalities: detection, recognition, localization, separation, enhancement, etc. Usually, these functionalities are handled by different 
sub-systems. However, we can expect that carrying out some of them jointly, i.e. with a single system and sharing a given processing framework, an advantage is obtained in terms of effectiveness and efficiency.

Often recognition and localization are required for the same application scenario. This may happen with speech 
(e.g. \cite{DIRHA1,6843271,6843273}), or with other types of acoustic events ($AE$) (e.g. \cite{Kotus2011,Goetze2012,ButkoCSGNHC11}). In this work, we have
developed a system that can detect, recognize and localize acoustic events, i.e. it estimates the identities of the acoustic events and their positions along the time axis and the spatial axes. 
The system carries out all those tasks jointly, employing a single processing scheme for all of them, a scheme that
uses models of the involved acoustic events. Furthermore, the system tackles the problem of simultaneity of acoustic events.

\begin{figure*}[!ht]
        \centering
        \includegraphics[width=0.85\textwidth]{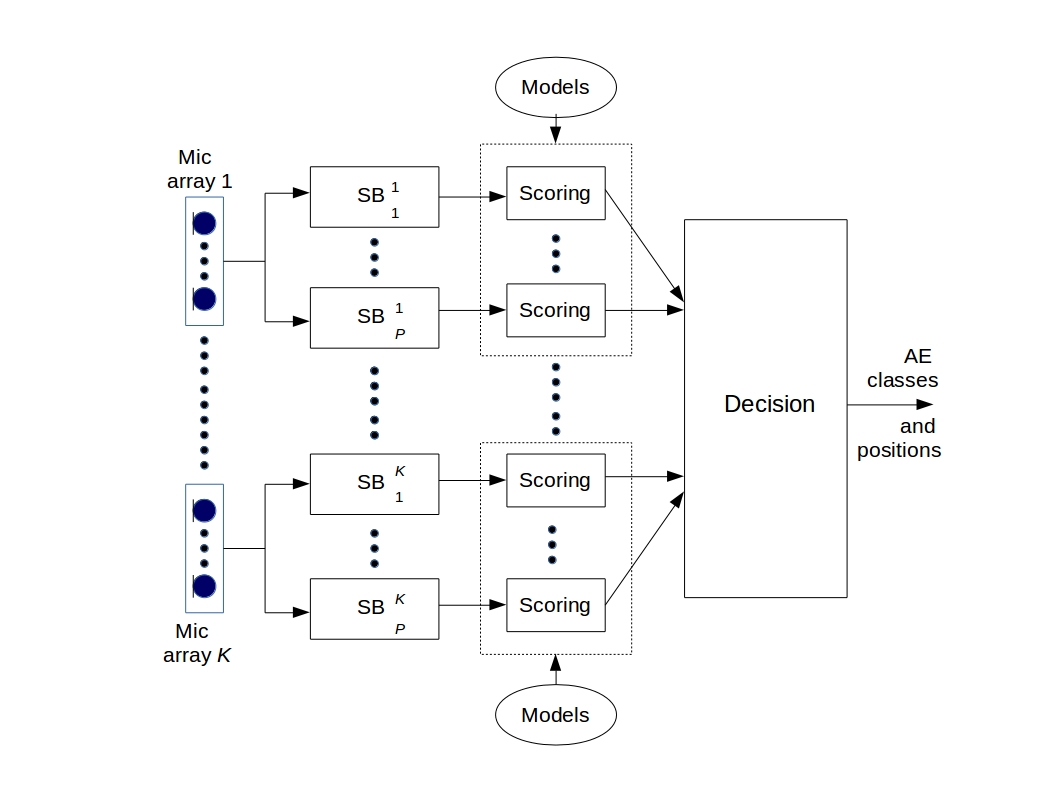}
        \caption{Processing scheme for both classification and localization. For each of the $K$ microphone arrays deployed in the room, a set of signals are produced at the output of $P$ array processors ($SB$).
Given an output signal, that results from targeting a specific room (cell) position, a score is
computed for each of the $C$ acoustic event classes using the set of models. The resulting $K$x$P$x$C$ scores are combined
with a given criterion to estimate both the class identity and the room position.}
        \label{fig:1}
 \end{figure*}
In fact, time overlap of acoustic events is a major factor of complexity for both acoustic event
recognition and acoustic source localization. Concerning the first, from the analysis of the results submitted by the participants 
in the meeting-room acoustic event detection ($AED$) task (which includes
recognition) of the $CLEAR’2007$ evaluation campaign \cite{Temko2009}, it was apparent that time overlapping of
acoustic events caused more than two thirds of the detection errors. More recently, in the $D$-$CASE$
evaluation \cite{giannoulis2013detection} a big gap was observed between the $AED$ accuracy of isolated and overlapped events. In
terms of $F$-score, the best results were $45.17$\% and $8.45$\%, respectively. In this paper, we deal with the
overlapping problem by taking into account the spatial diversity of the sound sources.

To estimate the position coordinates of the acoustic sources, most widely used acoustic source localization methods consider 
phase-based measures \cite{Brandstein2001Microphone,Dmochowski2010,Omologo93useof,4459286,citeulike:3783006}. Conversely, in this work we use the probability or similarity measure delivered by a classifier 
that employs the whole time-spectral information of the signals as input. As the classifier uses models for the different sound classes, those models can
be shared by both recognition and localization tasks.

In \cite{DBLP:conf/icassp/ChakrabortyN13}, a processing framework based on array beamformers and sound-model-based likelihood computations was introduced by these 
authors to recognize acoustic events that may overlap in 
time; there, recognition was actually reduced to a classification problem and only two microphone arrays were used. 
Posteriorly, a first trial for doing both recognition (actually, classification) and localization was reported in 
\cite{DBLP:conf/interspeech/ChakrabortyN13}, but only the direction of arrival was estimated and once the sound identity
was found. More recently, the same processing framework was applied to the source localization problem \cite{DBLP:conf/icassp/ChakrabortyN14}. 
In all those works, knowledge of the time end-points of the events was assumed in order to
compute event-level likelihoods. Conversely, that assumption is removed here. Furthermore, the
acoustic events, which may overlap, are jointly detected, identified and localized with a single system.
In other words, the presented system, which employs multiple arbitrarily-located small microphone
arrays, is able to determine the classes of the acoustic events (that occur either in isolation or simultaneously) 
along with their time end-points and their respective positions in the room space.

\begin{figure*}[!ht]
        \centering
        \includegraphics[width=0.9\textwidth]{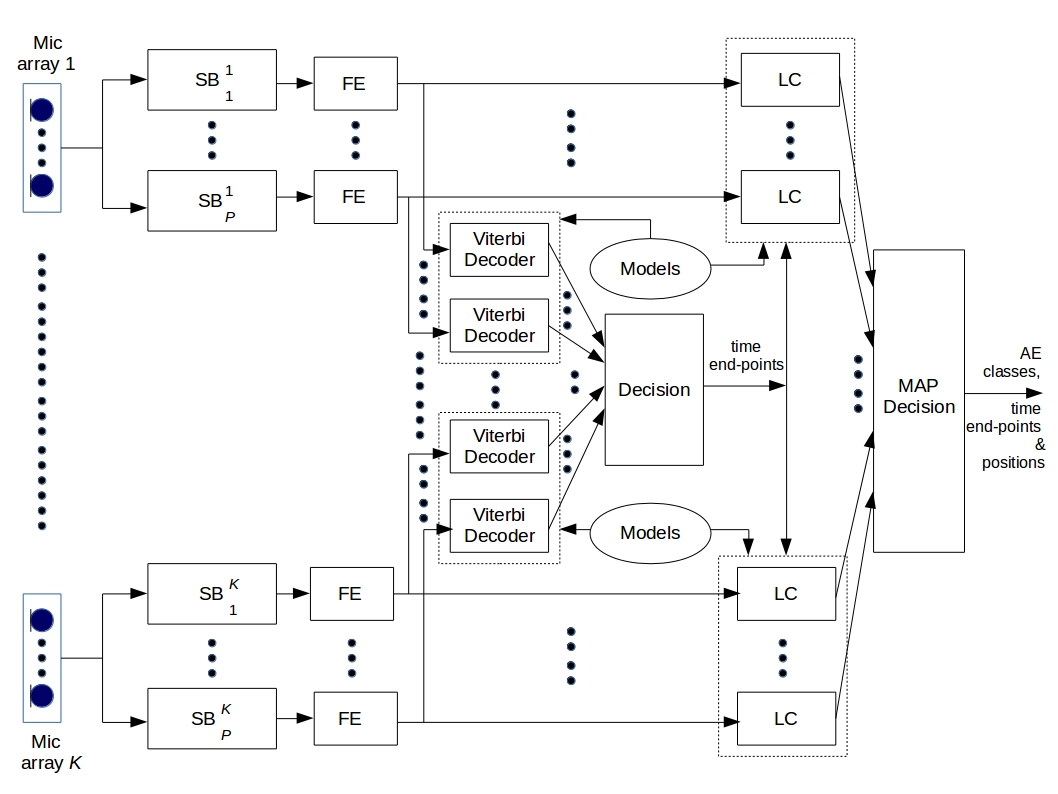}
        \caption{The proposed two-step recognition and localization system. The $Viterbi$ decoding of every channel has been 
        included as a first step. In the intermediate decision block the sequence of time end-points
that shows maximum likelihood is driven to the likelihood calculators from the second step. Note that steering beamformers (SB) and feature extractors (FE) are common to both steps.}
        \label{fig:4}
 \end{figure*}
Like in those previous works, the experiments reported in this paper are carried out for a concrete meeting-room scenario with 
one or two simultaneous sources, and using a database collected in the
smart-room, where six microphone arrays are distributed on the walls. For jointly doing recognition
and localization, a particular methodology is employed that follows a two-step approach. First the
events are detected along time, and then their class identities and space positions are estimated using a
maximum-a-posteriori criterion. The recognition and localization results are obtained with that two-step method using all 
the six three-microphone arrays in the room. Moreover, besides the experiments
with the section of the database that was employed in previous works, which consists of isolated sound
recordings, results with the other section of the database that includes actual two-source overlapping
sounds are reported. The proposed technique shows better recognition performance than a system
which models all the combinations of classes that occur in the database, and it shows better event-based space localization accuracy than the well-known $SRP$-$PHAT$ technique \cite{Omologo93useof}.

The rest of the paper is organized as follows. The conceptual approach is presented in Section \ref{sec:approach}, the developed system is described in Section \ref{Joint:RecogLoc}, and
in Section \ref{sec:experiments} we report the experimental work. Discussions and conclusions are given in Section \ref{sec:conclude}.

\section{Joint classification and localization approach}
\label{sec:approach}
Let us assume a room with a set of $K$ microphone arrays. Those arrays are arbitrarily located, so
they do not follow a spatially-structured configuration; for system deployment, this is an advantage.
The $2D$ room space is divided into a set of $P$ pre-defined small-area cells. Note that the vertical coordinate is not 
considered in this study, though it could be easily managed by the proposed system.

Along time, $N$ acoustic events take place, either in isolation or overlapped with other events. Those
events belong to a set of $C$ different acoustic classes. If we know the time end-points of every acoustic
event, we can jointly assign a class to it and estimate its spatial position with the processing scheme
depicted in Figure \ref{fig:1} already proposed by the authors in \cite{DBLP:conf/icassp/ChakrabortyN13,DBLP:conf/interspeech/ChakrabortyN13}. 
In fact, for each microphone array, a set of $P$ steering beamformers ($SB$) is designed, 
so that the $j^{th}$ beamformer, $1$$\leq$$j$$\leq$$P$, attenuates the
signals coming from all directions except the one that corresponds to position (cell) $s_{j}$. Then, each
beamformer output signal is separately fed into a classifier that computes a score for each pre-defined event class by using a set of $C$ pre-trained acoustic models. 
The whole set of $K$x$P$x$C$ likelihood scores is used by the decision block
to estimate both the identity and the source positions of the acoustic events using an optimality criterion.

The above processing scheme follows an event-based approach, i.e. to compute the likelihood with each class model, 
the time end-points of the (possibly overlapping) events must be known. In order to remove that constraint, 
i.e. to do detection instead of just classification, several different alternatives could be devised. In this paper, a two-step approach is taken: 
$(1)$ the events are firstly detected along time (but not recognized), and $(2)$ the identity and the source position of each event are determined. 
The first step outputs an estimated time interval (two time end-points) for each detected event, 
which is subsequently used in the second step for classifying and localizing.

\section{Joint recognition and localization system}
\label{Joint:RecogLoc}

The whole scheme of the two-step technique implemented in this work is depicted in Figure \ref{fig:4}. The
two steps share the initial part of the processing scheme, that includes beamforming and feature extraction. 
In the first step, a sequence of events is detected for each beamformer output signal. In our experimental work, 
the acoustic events are modeled with hidden Markov models ($HMM$), and the state
emission probabilities are computed with continuous density Gaussian mixture models ($GMM$). Consequently, 
after feature extraction, the sequence of observation vectors is decoded with the $Viterbi$ algorithm. 
Then, in the decision stage of this first step, the beamformer outputs are ranked according to
the likelihood given by the $Viterbi$ decoding. The time sequence of acoustic events corresponding to
the beamformer output that shows the highest likelihood is then taken to the second step. In fact, as
indicated in Figure \ref{fig:4}, only the time end-points of the hypothesized events are taken to the likelihood calculators of the 
second step, which use the same $HMM$-$GMM$ acoustic event models. If overlapped events are possible, 
the subsequent best sequences may also be taken to the second step.

In the second step, the last stage of the processing scheme from Figure \ref{fig:1} is applied for deciding the
class identities and the corresponding source positions of the acoustic events. Taking the time end-points of the events 
from the first step, and using the same set of statistical models that are used in that
first step, a $P$x$C$-dimensional matrix of likelihoods is build for each event and each microphone array.
Then, both the class identities and the space positions of the events are estimated with a $MAP$ criterion
and combining the probabilities from all arrays.

\subsection{Step 1: Time end-points estimation (detection)}
\label{step1}
Acoustic event recognition requires both segmentation of the audio stream and classification of the
segments. For a single beamformer output, simultaneous segmentation and classification can be performed like it is 
usually done for continuous speech recognition \cite{Rabiner:1993:FSR:153687,790984}. The goal for a single array can be
formulated as follows: given $X_{k}$, the set of output vectors from the $P$ feature extraction blocks corresponding 
to the $k^{th}$ array, find the event sequence $\Omega$ that maximizes the posterior probability $p(\Omega|X_{k})$:
\begin{equation}
 \hat{\Omega} =  \operatorname*{argmax}_{\Omega}p(\Omega|X_{k})  =  \operatorname*{argmax}_{\Omega}p(X_{k}|\Omega)p(\Omega)
 \label{eq1}
\end{equation}
where the likelihood $p(X_{k}|\Omega)$ is computed using statistical models, and $p(\Omega)$ is the prior probability of
the acoustic event sequence $\Omega$. In order to avoid the dependence of that sequence to the particular room
situation, all sequences of events are assumed equally probable, i.e. $p(\Omega)$ is constant and it does not affect
the maximization in Equation \ref{eq1}.

As shown in Figure \ref{fig:4}, given a set of $P$ beamformers for each of the $K$ microphone arrays, the decoding
of every channel with the $Viterbi$ algorithm \cite{Rabiner:1993:FSR:153687,790984} yields a set of $L$=$P$x$K$ hypothesis, each consisting of a sequence of detected 
events with their corresponding end-points and its likelihood score. In the decision
stage, the best channel is chosen by maximizing the posterior probability $p(\Omega_{l}|X_{k})$, $1$$\leq$$l$$\leq$$L$ across all channels.

The end-points of the acoustic events corresponding to the best channel are used in the second step
for jointly doing classification and localization with the processing scheme depicted in Figure \ref{fig:1}. Note
that, although the identities of the detected sounds are also hypothesized by this first-step algorithm,
they are not used in the $2^{nd}$ step since a better accuracy is obtained using the MAP criterion in that step.
Also, though an operationally simplified system would result from using just one sound class model
that encompass all events, the end-points accuracy would suffer from the lack of specificity of that
overall sound model, and model training would be more demanding as the specific $AE$ models are also
needed in $2^{nd}$ step but the single overall model would have to be additionally created.

\subsection{Step 2: MAP-based classification and localization}
\label{step2}
To decide about the class identities and positions, all the channels (from all the arrays) are taken into account. 
As depicted in Figure \ref{fig:4}, the feature vector sequence (already extracted in $Step$ $1$) from each
beamformer output signal enters the classification system. Then, a likelihood score ($LC$) is computed
for each of the event time intervals from $Step$ $1$ by using the same $HMM$-$GMM$ models that were used
in that $Step$ $1$. Finally, a decision module carries out the classification and localization of the events by
combining likelihood scores and priors using a $MAP$ criterion and combining the probabilities from all arrays.

Given a room with $K$ microphone arrays, let us assume we have a set of $N$ events with labels $c_{i}$,
$1$$\leq$$i$$\leq$$N$, which belong to a set of $C$ different classes. Given a grid of room positions $s_{j}$, 
$1$$\leq$$j$$\leq$$P$, for each array, there is a set of $P$ beamformers, so that the $j^{th}$ beamformer attenuates the signals that come
from all positions except position $s_{j}$. Thus, from that initial array signal processing block, we have a set
of $P$ output signals for each array, and after feature extraction and likelihood computations with the
models of all classes, we have a $K$x$P$x$C$-dimensional vector of likelihood scores. For identifying the
sound class and localizing the sound sources, we first determine the posterior probability of a given
class $c_{i}$ and position $s_{j}$ for each $k^{th}$ array, i.e.
\begin{equation}
 p(c_{i},s_{j}|X_{k}) =  p(X_{k}| c_{i},s_{j})p(c_{i})p(s_{j})/p(X_{k})
 \label{eq2}
\end{equation}
where $p(X_{k}| c_{i},s_{j})$ is the likelihood of $X_{k}$ given class $i$ and position $j$, $p(c_{i})$ is the (prior) probability of 
class $i$, $p(s_{j})$ is the (prior) probability of position $j$, and $p(X_{k})$ is the probability of the feature vector $X_{k}$.

In order to combine the posterior probabilities from the various microphone arrays, the product
combination rule is used \cite{Kuncheva:2004:CPC:975251}, and so the optimal class $c_{o}$ and the optimal position $s_{o}$ 
are chosen so as to maximize the product of posterior probabilities, i.e.
\begin{equation}
 c_{o},s_{o} =  \operatorname*{argmax}_{c_{i},s_{j}}\prod_{k=1}^{K}p(c_{i},s_{j}|X_{k})
 \label{eq3}
\end{equation}
Note that for finding the maximum-a-posteriori (MAP) estimates $c_{o}$ and $s_{o}$, the probability of the feature vector $p(X_{k})$ is not required.
Moreover, both the priors of classes and positions can be included in the optimization. In the experiments we will consider the latter.


\subsection{Feature extraction and modeling}
\label{sec:AEDASL}
\begin{figure*}[!ht]
        \centering
        \includegraphics[width=0.8\textwidth]{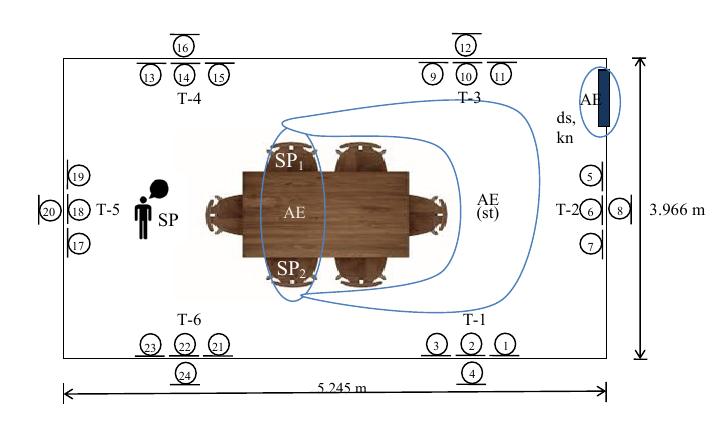}
        \caption{Smart-room layout, with the positions of microphone arrays ($T$-$k$), acoustic events ($AE$),
speaker ($SP$) in $S$-recordings, and speakers ($SP$ $1$ and $SP$ $2$) in $T$-recordings. Some place-specific acoustic
events are indicated in the plot: steps ($st$), door-slam ($ds$) and door-knock ($kn$)}
        \label{fig:5}
 \end{figure*}
In the feature extraction block of the system, a set of audio spectro-temporal features is computed for each signal frame. 
Frames are $30$ $ms$ long, with a $20$ $ms$ shift, and a $Hamming$ window is applied. We have used frequency-filtered logarithmic filter-bank energies ($FF$-$LFBE$) for the parametric representation of the 
spectral envelope of the audio signal, as it generally is a more robust alternative to
$MFCC$ features and requires a smaller amount of computations than them \cite{DBLP:journals/speech/NadeuMH01}. For computing the $FF$-$LFBE$s 
in a given frame, the $FIR$ filter with transfer function $z$-$z^{-1}$ is applied to the sequence of log sub-band energies. This is equivalent to a convolution in the frequency domain that actually computes a smoothed derivative along frequency. 
Zeros are assumed beyond the vector ends and, after convolution, the end-points are also taken into account. 
In our experiments, $16$ sub-bands are considered, so obtaining $16$ $FF$-$LFBE$ features per frame. 
Additionally, as it is usually done in speech recognition, temporal smoothed derivatives are also used to somewhat 
include the temporal evolution of the envelope. Therefore, the final feature vector, which has $32$ elements, 
contains the first-order smoothed derivatives of the log sub-bank energies both along frequency and time.

Hidden Markov Models are used for modeling the $AE$s (including the speech event), and their state emission probabilities are
modeled with continuous density Gaussian mixture Models. There is one left-to-right $HMM$ with three
emitting states for each $AE$ and for each array. $32$ Gaussian components with diagonal covariance matrix are used per state. 
Each $HMM$-$GMM$ model is trained with the standard $Baum$-$Welch$ algorithm \cite{Rabiner:1993:FSR:153687},
using all the signals from an array with the $HTK$ toolkit \cite{htkbook}. The same models are used for both
processing steps mentioned in Sub-section \ref{Joint:RecogLoc}. To avoid mismatch between training and testing, the
signals at the output of the beamformers are used for training the models. If each beamformer had its
own set of models (for all acoustic events and silence), the total number of models would be too large.
That is why we chose to have a common set of models for all the beamformers in a given array.

In the second step, where the optimal class identity and source position are obtained as indicated by
Equation \ref{eq3}, flat priors are assigned to classes in the reported tests. However, regarding the positions,
either flat or estimated prior probabilities are used. The latter are estimated by counting the event occurrences within 
the training dataset and computing the quotient between the number of occurrences in
the cell and the total number of event occurrences in the dataset.

\section{Experimental work}
\label{sec:experiments}
In the experimental work, a meeting-room scenario with a predefined set of eleven acoustic events
plus speech is considered \cite{Temko2009}. Like in \cite{ButkoCSGNHC11}, we assume that there may occur either $0$, $1$ or
$2$ simultaneous events.
\subsection{Meeting-room acoustic scenario and database}
\label{sec:database}
Figure \ref{fig:5} shows the smart-room layout with the position of its six $T$-shaped four-microphone arrays on
the walls \cite{DBLP:journals/ejasmp/ChakrabortyNB14}. 
The steering beamformers at the front-end of the system are designed to work with the horizontal row of three microphones available for each of the
For system training, development and testing, we have used the audio part of a multi-modal database 
recorded in that smart-room with all microphones \cite{ButkoCSGNHC11} \footnote{That database is publicly available from the authors.}. 
There are $11$ meeting-room acoustic event ($AE$) classes:  applause, spoon$/$cup jingle, chair moving, cough, door open$/$slam, key jingle, 
door knock, keyboard typing, phone ringing, paper work, and steps. The regions where the $AE$ sources are located are shown
in Figure \ref{fig:5}. Signals were recorded with all the six $T$-shaped microphone arrays ($24$ microphones in total),
but only the signals coming from the three horizontally-aligned microphones from every array are used in the experiments.

The database contains two different sections or datasets, and both will be used in these reported experiments. 
The first one, so called $S$-recordings, consists of $8$ audio recording sessions which contain isolated acoustic events. 
For training, only these one-source signals are used in our experiments.
The audio data corresponding to two simultaneous sources that are used for development and testing were obtained by 
adding those $AE$ signals recorded in the room with a speech signal also recorded in
the room from all $24$ microphones. To do that, for each $AE$ instance, a segment with the same length
was extracted from the speech signal starting from a random position, and added to the $AE$ signal. The
mean power of speech was made equivalent to the mean power of the overlapping $AE$. That addition of
signals produces an increment of the background noise level, since it is included twice in the overlapped signals; however, going from isolated to overlapped signals the $SNR$ reduction is slight: from
$18.7$ $dB$ to $17.5$ $dB$. Although in our real meeting-room scenario the speaker may be placed at any point
in the room, its position in the $S$-recordings dataset is fixed at a point at the left side (‘$SP$’ in Figure \ref{fig:5}).
However, this will not constrain the usefulness of the results, because the proposed system will not
make use of that knowledge.

Additionally, a more realistic dataset was also used for testing: the $T$-recordings. In it, two participants 
(room positions ‘$SP$ $1$’ and ‘$SP$ $2$’ in Figure \ref{fig:5}) interact with each other in a rather spontaneous way.
While one is speaking, the other is randomly producing acoustic events, which belong to the above
mentioned set of $11$ meeting-room $AE$ classes. Therefore, this dataset includes overlaps of an $AE$ with
speech that are more naturally produced than in the $S$-recordings dataset. All signals were recorded at
$44.1$ $kHz$ sampling frequency, and they were further down-sampled to $16$ $kHz$ for our experiments.

\subsection{Experiments and results}
\label{sec:Results}
In the reported experiments, the proposed system, depicted in Figure \ref{fig:4}, is used to recognize the $AE$
and to localize either one or two simultaneous acoustic event sources in the room environment.
\begin{figure}
        \centering
        \includegraphics[width=0.45\textwidth]{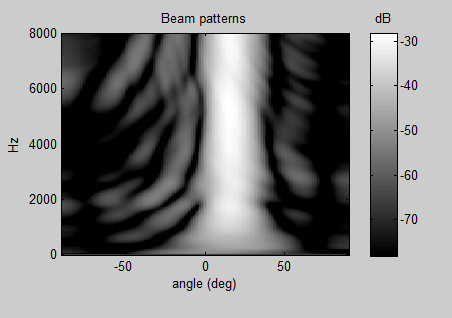}
        \caption{Illustration of a frequency-invariant beam pattern}
        \label{fig:3}
 \end{figure}
\begin{table*}[]
\centering
\caption{Classification accuracies (\%) for the various systems in the two-source scenario}
\label{Table:1}
\begin{tabular}{|c|c|c|c|c|c|}
\hline
            &                                                                  &      & \multicolumn{2}{c|}{Proposed technique (1-step)} &                                                          \\ \hline \hline
            & \begin{tabular}[c]{@{}c@{}}All-combinations\\ modeling\end{tabular} & BSS  & Flat priors          & Estimated priors          & \begin{tabular}[c]{@{}c@{}}Known\\ position\end{tabular} \\ \hline \hline
$S$-recordings & 83.8                                                             & 80.8 & 84.4                 & 86.7                      & 89.1                                                     \\ \hline \hline
$T$-recordings & 68.2                                                             & 67   & 83.5                 & 85.2                      & 83.7                                                     \\ \hline
\end{tabular}
\end{table*}
\begin{table*}[]
\centering
\caption{Recognition accuracies (AED-ACC in \%) in the two-source scenario}
\label{Table:2}
\begin{tabular}{|c|c|c|c|}
\hline
             & \begin{tabular}[c]{@{}c@{}}All-combinations\\ modeling\end{tabular} & BSS  & \begin{tabular}[c]{@{}c@{}}Proposed technique,\\ only first step\end{tabular} \\ \hline \hline
$S$-recordings & 80.1                                                              & 77.6 & 80.4                                                                          \\ \hline \hline
$T$-recordings & 65.6                                                              & 63.8 & 75.6                                                                          \\ \hline 
\end{tabular}
\end{table*}
In order to deal with the broadband characteristics of the audio signal 
a frequency invariant beamforming used to determine the beamformer coefficients. The method,
proposed in \cite{:/content/asa/journal/jasa/119/6/10.1121/1.2197606}, uses a numerical approach to construct 
an optimal frequency invariant response for an arbitrary spacing a small number of microphones. 
It first decouples the spatial selectivity from the frequency selectivity by replacing the set of real sensors
by a set of virtual ones, which are frequency invariant. Then, the same array coefficients can be used
for all frequencies.

All six microphone arrays ($T1$ to $T6$) available in the room are employed in the experiments. The steering beamformers at the front-end of the system are designed to work with the horizontal row of three microphones available for each of the
arrays in the smart-room. An illustrative example of the beam pattern is shown in Figure \ref{fig:3} \cite{DBLP:journals/ejasmp/ChakrabortyNB14}. Note how the beams
for the angle of interest are rather constant along frequency. Notice also that, due to the low number of microphones used in this particular experimental setup, the beamformer is steering to the target cell with a relatively broad beam. Regarding the design of the room grid considered in the experiments, as high localization precision is not required for the meeting-room application, a rather large cell size of $0.661$ $m$ x $0.874$ $m$ has been chosen, and so there are $36$ cells ($6$x$6$).

Testing results are obtained for both types of recordings and for all the sessions (i.e. eight sessions
$S01$ to $S08$ for $S$-recordings, and nine sessions, $T01$ to $T09$, for $T$-recordings), using a leave-one-out
criterion, i.e. recursively keeping one session for testing, while all the other sessions are used for training. 
The number of simultaneous sources is assumed known ($1$ or $2$). For the two-source case, as mentioned
above, one sound is always speech and the other is any one of the eleven acoustic events. However, the
proposed system does not use that knowledge. In that two-source case, the identities and source positions of the two overlapping 
sounds are found by applying Equation \ref{eq3} two consecutive times and leaving out
the recognized class and the estimated position after the first time. Although speech is modeled as another 
acoustic event and recognized, only the system outputs corresponding to the eleven $AE$s are considered for computing the evaluation scores.

\begin{table*}[]
\centering
\caption{Recognition accuracies (AED-ACC in \%) for the two-source scenario}
\label{Table:3}
\begin{tabular}{|c|c|c|c|}
\hline
             & \multicolumn{2}{c|}{Proposed technique (2-steps)} &                 \\ \hline 
             & Flat priors           & Estimated priors          & Known positions \\ \hline \hline
$S$-recordings & 82.3                  & 84.2                      & 87.2            \\ \hline \hline
$T$-recordings & 77.8                  & 79.1                      & 78.7            \\ \hline
\end{tabular}
\end{table*}
To evaluate both the recognition and the localization performance of our system, a $F$-score metric
(harmonic mean between precision and recall) is used. That metric is defined as the one that was used
for event detection in the $CLEAR'2007$ \cite{Temko2009} international evaluation.

\subsubsection{Classification experiments}
\label{sec:CE}
All the classification experiments are carried out for the two-source scenario. First of all, we present
a test that was done assuming that the time end-points of the events are known, so it really consists in a
classification (not detection) experiment. In that test, the $AE$ classification performance of the proposed system 
from Figure \ref{fig:1} (one-step system) is compared with those from two reference systems, which
use only one channel (microphone $18^{th}$). The first reference system uses a model for every possible
combination of simultaneous events (all-combinations modeling). That is, the system has a model for each possible class whether it is 
an isolated event or a combination of events \cite{Temko:2009:AED:1595906.1596207}. In our
two-source scenario, this approach requires $11$ models for the isolated $AE$s plus $11$ models for the $AE$s
overlapped with speech. Therefore, it does not need a prior separation of the two overlapped signals,
but requires a number of models that may be too large in applications with a high number of classes or
a high degree of simultaneity. Tests are carried out with a HMM-GMM based system \cite{DBLP:conf/eusipco/ButkoPSNH11} 
that was developed and tested with the same (S-recordings) database and accuracy metric used in this paper. The second reference system uses a blind source separation ($BSS$) approach that consists of a 
deflation-based iterative technique, where the source signals are extracted
from the mixtures one by one \cite{Simon2001883,4034126}. To reduce the time complexity for this $BSS$ technique, a quadratic contrast 
function with $4^{th}$ order cumulants is used \cite{4034126,4960295}. In our experiments, we avoid the permutation problem of the algorithm 
by doing the right choice at the output of the separation block, before the classifier.
\begin{table*}[]
\centering
\caption{Average localization accuracy (in \%), assuming the time end-points of events are known}
\label{Table:4}
\begin{tabular}{|c|c|c|c|c|}
\hline
                              &          &  $SRP-PHAT$ & \multicolumn{2}{|c|}{Proposed technique} \\ \hline 
                              &          &  & Flat priors     & Estimated priors     \\ \hline \hline
\multirow{2}{*}{$S$-recordings} & 1 source & 86.9     & 86.4            & 88.1                 \\ \cline{2-5} 
                              & 2 source & 70.9     & 83.9            & 85                   \\ \hline \hline
$T$-recordings                  &          & 82.3     & 85.6            & 86.2                 \\ \hline
\end{tabular}
\end{table*}

The acoustic event classification results are shown in Table \ref{Table:1}. For this experiment, the performance
of the various systems is measured in terms of classification accuracy, which is defined as the quotient
between the number of correctly classified events and the total number of occurrences in the testing
database. The results for $S$- and $T$-recordings are shown in two different rows in the table. Regarding
the proposed joint system, two cases are considered using for the source
positions either the flat or the estimated priors of source positions. The results for the $S$-recordings are also compared with
those from a (virtual) version of the proposed system that assumes the source positions are known.

Notice from Table \ref{Table:1} that the proposed system shows a higher classification accuracy than the other
two systems (all-combinations modeling and $BSS$-based), though a lower one than the system that
assumes the source positions are known. As expected, the use of estimated position priors produces
better classification accuracy for the proposed system than the use of flat priors. 
Logically, the classification accuracies are higher for the somewhat artificially-mixed signals from the $S$-recordings. When
using the more realistic $T$-recordings, the performances from the two reference systems suffer a much
significant degradation than the ones of the proposed technique. Concretely, the percentage of classification error 
reduction of the proposed technique with respect to the all-combinations modeling technique for the case of estimated priors is as high as 53.5\%.
Notice that, with estimated priors, the
proposed system offers for $T$-recordings a higher accuracy than the system that uses knowledge of
source positions. This apparent contradiction can be attributed to the fact that the experiment with
known positions was carried out assuming that during recordings the two participants did not move from the room position initially 
assigned to each of them, but they actually moved due to the spontaneity requirement.

\subsubsection{Recognition experiments}
\label{sec:RE}
In the following, we report the recognition results, i.e. the ones obtained when the time end-points of the events are
assumed unknown. As in Sub-section \ref{sec:CE}, all the experiments are carried out for the two-source scenario, but
only the (non-speech) acoustic event is evaluated. First of all, only the first step of the proposed algorithm is tested to check how it is working.
Its performance, which is measured in terms of the $F$-score (the $AED$-$ACC$ metric \cite{Temko2009}), computed from the $Viterbi$-
decoding-based best channel is compared with those from the previously reported systems (all-combinations modeling and $BSS$-based). 
Notice from Table \ref{Table:2} that the proposed one-step recognition technique outperforms the $BSS$-based one and gets a similar result 
to that from the all-combinations modeling technique. Again, the degradation of the proposed algorithm is smaller than the
ones from the other two techniques when moving to the more realistic $T$-recordings.

Table \ref{Table:3} shows the recognition results for the proposed two-step system, which can jointly estimate
the $AE$ identity and its cell-based position, using either flat or estimated priors of source positions. Like for the classification results 
in Table \ref{Table:1}, here the performance of a reference system that assumes known source positions is also presented as reference. 
By comparing with the last column in Table \ref{Table:2}, notice that the addition of the
second step increases the recognition performance, especially when estimated priors are employed.
That improvement is due to the fact that the two-step algorithm is effectively combining the likelihood
scores from all the channels instead of choosing only the channel that gets the highest overall likelihood computed with the $Viterbi$ algorithm. 
Again, the results from the $T$-recordings are worse than those from the $S$-recordings, but the proposed system shows a lower degradation. 
Actually, the error reduction percentage with respect to the all-combinations modeling technique for the $T$-recordings is 39.2\% 
when using estimated priors. Again, the lower score for the known positions case can be explained by the inadequacy of the 
assumption about steady participant position in those recordings.
As it can be expected, a significant degree of degradation is observed going from known to unknown 
end-points, i.e. by including the automatic detection in the first step. 
In fact, by comparing Table 1 with Table 3 we notice that the loss in terms of relative increase of error from 
classification to recognition when estimated priors are used is 18.8\% for S-recordings and 41.2\% for T-recordings.

\subsubsection{Localization experiments}
\label{sec:LE}
Acoustic source localization is carried out at the event level for all the tested techniques, and using
either one-source or two-source signals. As we did with the presentation of the recognition results,
we firstly assume that the time end-points of the events are known. Consequently with that, the metric
used is the localization accuracy, which is defined for a given class as the quotient between the number of correct localizations 
and the total number of event occurrences in the testing dataset. For a given
event occurrence, a correct localization happens when the cell assigned to the true position is the same as the one estimated by the system. 
The true position for each event was obtained from visual inspection during signal recordings.

For the localization tests, the proposed system is compared with a reference $SRP$-$PHAT$ system \cite{Omologo93useof,ChaZhang,conf/icassp/DoSY07}. 
The $SRP-PHAT$ localization technique explores the room space searching for the maximum
of the global contribution of the $PHAT$-weighted cross-correlations from all the microphone pairs. To
efficiently find the global maximum, the stochastic region contraction algorithm is used \cite{conf/icassp/DoSY07,97993} in our
tests. $SRP$-$PHAT$ works at the frame level so, in order to evaluate its results at the event level, we have
averaged the estimated position coordinates along the whole event time interval (averaging gave us better results than the use of a voting procedure), which is assumed known.

The results for the case of known time end-points are presented in Table \ref{Table:4} in terms of average of accuracies along $AE$ classes. 
As can be observed in that table, the joint recognition and localization system (the proposed technique) performs
better in terms of localization accuracy than the $SRP$-$PHAT$ system, both for $S$- and $T$-recordings. 
Notice again that the degradation from the one-source
case to the two-source one is much smaller for the proposed technique. 
The error reduction in the two-source case with respect to $SRP$-$PHAT$ is 48.4\% for estimated priors, in
spite that $SRP$-$PHAT$ was implemented by looking at the two maxima of the sound map and counting a correct localization 
if anyone of the maxima was in the correct cell. Presumably, the inclusion of
the time-spectral content of the $AE$s in the proposed system
helps to better localize the sources, especially in the more complex two-source scenario.
\begin{table}[]
\centering
\caption{Average localization accuracy (in \%) of the proposed joint approach}
\label{Table:5}
\begin{tabular}{|c|c|c|c|}
\hline
                              &          & \multicolumn{2}{c|}{Proposed technique} \\ \hline 
                              &          & Flat Priors      & Estimated priors     \\ \hline \hline
\multirow{2}{*}{$S$-recordings} & 1 source & 85.5             & 87.3                 \\ \cline{2-4} 
                              & 2 source & 83.6             & 84.6                 \\ \hline \hline
$T$-recordings                  &          & 84.7             & 85.4                 \\ \hline
\end{tabular}
\end{table}
%
Table \ref{Table:5} shows the localization performance of the proposed system when the time end-points of the events are unknown 
and so they are estimated in the first step of the system. The performance is measured in terms of the $F$-score metric mentioned 
in Sub-section \ref{sec:Results}. The resulting scores are only slightly worse than those from the experiment reported in Table \ref{Table:4} 
with the system that uses knowledge of time end-points. Actually, the relative error increments are 6.7\% for one source, 
2.7\% for two sources, and 5.8\% for T-recordings. It is worth noticing that, in terms of localization performance, the system is 
less prone to end-point estimation errors than in terms of recognition performance.
\section{Conclusions}
\label{sec:conclude}
A novel model-based approach for joint recognition and localization of simultaneously-occurring
meeting-room acoustic events has been presented in this paper. Using multiple arbitrarily-located small
microphone arrays and $HMM$-$GMM$ models, a specific two-step system has been developed and tested. 
From the experiments, we observe that the system yields remarkably higher recognition scores than two common techniques, 
especially for the more realistic $T$-recordings. Although those two
techniques do not require multichannel signals, their computational processing is not shared by a localization system as in the proposed 
technique, and furthermore, if all the classes can appear in all sources, the number of models required by the all-combinations modeling 
technique may be very high, since it is $C^{S}$, being $S$ the number of simultaneous sources.
Also, the event-based localization performance of the proposed system 
is significantly better better than that of the widely used $SRP$-$PHAT$ method, especially in the two-source scenario,
although the latter technique is able to work at the finer signal segment level. 
Actually, the proposed technique is more computationally expensive than the $SRP$-$PHAT$ technique, but the required computations are shared with the recognition system. 
In conclusion, the proposed technique can be a proper alternative for localization in a multiple source scenario when it works together 
with an audio recognition system, since besides not requiring extra computations, it takes advantage of the sound characteristics in a multiple source scenario.
Finally, the system is able to use prior probabilities of event
class and room position, and we observed that both the recognition and the localization performance
improve significantly by using the estimated priors of room positions. A further improvement can be expected also from using the event class probabilities whenever
they are not flat. Notice that, since event sources will generally have different probabilities of occurrence at different room positions, 
the use of a joint class-position prior could further increase the system performance. In summary, the proposed approach shows the
advantage of carrying out the two tasks, recognition and localization, with a single model-based system.
\section{Acknowledgment}
\label{sec:ackw}
This work has been supported by the Spanish projects $SARAI$ ($TEC2010-21040-C02-01$) and $SPEECHTECH4ALL$ ($TEC2012-38939-C03-02$).
\newpage
\bibliographystyle{IEEEbib}

  \bibliography{Sensors_TALP_CNRC.bib}

\end{document}